\documentclass[9pt,conference]{IEEEtran}
\usepackage{waspaa25} % removes hyperlinks (required by IEEE Xplore)

\usepackage{bm} % for bold math symbols (incl. Greek letters) with \bm{}

\title{IS³ : Generic Impulsive--Stationary Sound Separation \\ in Acoustic Scenes using Deep Filtering}

\name{Clémentine Berger$^{1}$,
      Paraskevas Stamatiadis$^{1}$,
      Roland Badeau$^{1}$,
      Slim Essid$^{1}$ \thanks{This work was supported by the Audible project funded by French BPI, and was performed using HPC resources from GENCI-IDRIS (Grant 2024-AD011014883R1). 
}}
\address{$^{1}$LTCI, Télécom Paris, Institut Polytechnique de Paris, France}

\begin{document}

\maketitle

\begin{abstract}
We are interested in audio systems capable of performing a differentiated processing of stationary backgrounds and isolated acoustic events within an acoustic scene, whether for applying specific processing methods to each part or for focusing solely on one while ignoring the other. 
Such systems have applications in real-world scenarios, including robust adaptive audio rendering systems (e.g., EQ or compression), plosive attenuation in voice mixing, noise suppression or reduction, robust acoustic event classification or even bioacoustics.
To this end, we introduce IS³, a neural network designed for Impulsive--Stationary Sound Separation, that isolates impulsive acoustic events from the stationary background using a deep filtering approach, that can act as a pre-processing stage for the above-mentioned tasks. To ensure optimal training, we propose a sophisticated data generation pipeline that curates and adapts existing datasets for this task. We demonstrate that a learning-based approach, build on a relatively lightweight neural architecture and trained with well-designed and varied data, is successful in this previously unaddressed task, outperforming the Harmonic--Percussive Sound Separation masking method, adapted from music signal processing research, and wavelet filtering on objective separation metrics.
\end{abstract}

\section{Introduction}
\label{sec:intro}
An acoustic scene can be roughly decomposed into a stationary ambient background, containing a mixture of environmental sounds (wind, rain, insects etc.) and anthropogenic noises (traffic noise, speech babble noise or murmur, ventilation noise etc.), overlayed with isolated and \textit{impulsive} acoustic events. These impulsive events are characterized by a sudden increase in sound pressure level over a short duration and can stand out to varying degrees from the background. Examples include impacts, explosions, bursts, clapping, short alarms, or even coughing...
In many contexts, these two categories of sounds, stationary ambient backgrounds and impulsive events, require distinct and independent processing due to their differing characteristics. This is particularly relevant in audio mixing (e.g., differentiated equalization and compression) or audio pre-processing for tasks such as speech enhancement, and noise reduction/suppression.

\noindent \textbf{Related works.} \
To enable such a differentiated processing, separating the stationary background from impulsive sounds may be beneficial, allowing for targeted treatments. However, this specific separation task remains under-explored in the literature. 
Existing approaches primarily focus on impulsive noise attenuation or suppression for specific applications such as music restoration \cite{brajevic2011elimination, ruhland2015reduction}, speech communication \cite{godsill1996robust, subramanya2007automatic, sugiyama2007single, nongpiur2008impulse, sugiyama2013impact}, and specialized domains such as automotive or aeronautical noise reduction \cite{medda2017separation, wodecki2019impulsive} and bioacoustics \cite{young2013}.
These methods often target context-specific noise (including audio artefacts rather than distinct sound events), and rely mostly 
on signal processing techniques that first detect impulsive events and subsequently remove them using interpolation and magnitude adjustment techniques \cite{sugiyama2013impact} or separate them through reconstruction methods \cite{wen2004separation}, masking \cite{ruhland2015reduction}, or wavelet filtering \cite{nongpiur2008impulse, brajevic2011elimination, medda2017separation}.

In contrast, we focus on general acoustic scenes from everyday life, aiming to separate and reconstruct both ambient backgrounds and the impulsive sounds as faithfully as possible to support downstream applications.  
Blind Source Separation (BSS) methods appear well-suited for this task. Some studies have explored matrix demixing using statistical signal analysis \cite{sahmoudi2005blind}, while others have focused on time-frequency (TF) domain masking approaches. Notably, in the musical domain, the Harmonic-Percussive Source Separation (HPSS) method \cite{fitzgerald2010harmonic} has been proposed, leveraging median filtering along both the time and frequency axes to generate harmonic and percussive masks for source separation. 

More recently, deep learning approaches, particularly deep filtering, have surpassed traditional ratio-masking for speech enhancement \cite{mack2019deep}, estimating complex-valued time-frequency filters that captures correlations with adjacent TF bins and improving the extraction process.
However, this comes at the cost of increased computational complexity. To address this, DeepFilterNet \cite{schroter2022deepfilternet, schroter2022deepfilternet2} balances deep filtering and real-gain predictions on an equivalent rectangular bandwidth (ERB) spectral representation, achieving state-of-the-art performance while remaining lightweight for real-time applications.

\noindent \textbf{Contributions.} \
We propose IS³, a deep filtering approach for Impulsive--Stationary Sound Separation in ambient acoustic scenes, aimed at reconstructing both impulsive components and the stationary background. A key challenge in this task is obtaining high-quality training data for supervised learning, which requires a diverse set of clean acoustic scenes free from impulsive sounds, combined with an extensive variety of isolated impulsive sounds.
Our contributions are: %\\
i) a methodology for curating and adapting existing datasets to this task, along with a procedure for generating training, validation, and test data by combining these datasets; %\\
ii) a learning-based approach for the task of impulsive--stationary sound separation build on the adaptation of the DeepFilterNet architecture \cite{schroter2022deepfilternet2}; %\\
iii) an extensive evaluation on realistic data showing the superiority of our system to previous approaches including the HPSS masking method and an adaptation of the wavelet-based process from Nongpiur's article \cite{nongpiur2008impulse}.
\section{Model}

We first provide an overview of our system IS³, followed by a description of the loss functions used for optimization.
\begin{figure*}[ht]
    \centering
    \includegraphics[width=0.95\textwidth]{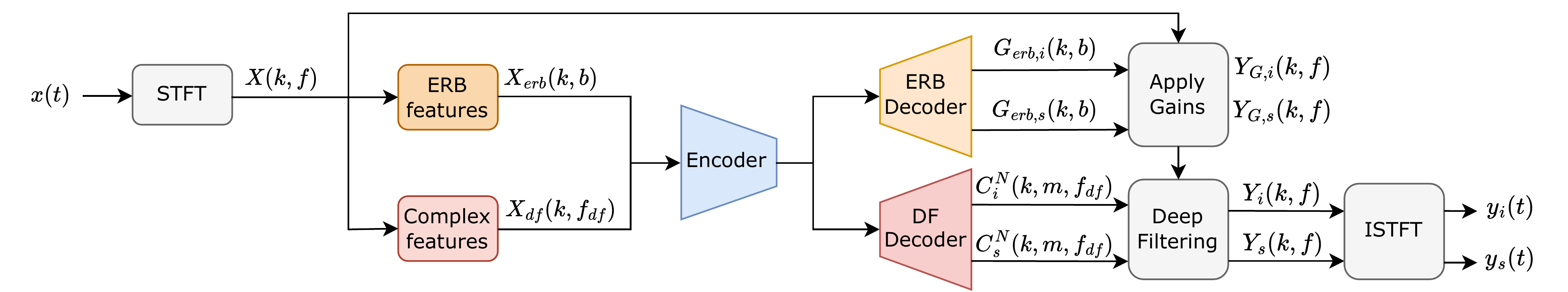}
    \caption{Overview of the IS³ system: The model extracts real ERB and complex features from the input mixture and processes them through a shared encoder. Two decoders then operate in parallel: one predicts real-valued ERB gains for impulsive and stationary components, while the other estimates complex time-frequency filters for each source. These gains and filters are applied to the input spectrogram, and the separated signals are reconstructed via inverse STFT.}
    \label{fig:system_overview}
\end{figure*}

\subsection{System overview}
%The method used for the impulsive/stationary separation task is based on deep filtering \cite{mack2019deep} which consists in the prediction of complex time-frequency filters. 
The architecture chosen for IS³ is strongly inspired by that of DeepFilterNet \cite{schroter2022deepfilternet, schroter2022deepfilternet2} for speech enhancement, adapted here for impulsive--stationary sound separation as presented in \autoref{fig:system_overview}. The model follows an encoder-decoder structure that predicts parameters for a two-stage filtering process, corresponding to varying levels of filtering precision. The first stage predicts real-valued gains defined on ERB frequency bands, while the second stage performs deep filtering (DF) by predicting a complex filter.

The IS³ system takes as input an acoustic signal $x(t)$ sampled at 44100 Hz, which we decompose as follows:
\begin{equation}
    x(t) = y_i(t) + y_s(t),
\end{equation}
where $y_s$ represents the stationary part of the acoustic scene and $y_i$ denotes the impulsive components.
The separation process operates in the frequency domain:
\begin{equation}
    X(k,f) = Y_i(k,f) + Y_s(k,f),
\end{equation}
where $X$ is the short-time fourier transform (STFT) of $x$, computed with an analysis window of $N_{fft} = 2048$ with a 75\% overlap, and a Hanning window. The indices $k$ and $f$ represent time and frequency bins, respectively.

The encoder processes both magnitude and complex features, as described in \cite{schroter2022deepfilternet}. Magnitude features, denoted by $X_{erb}(k,b), b \in [1, N_{erb}]$, are extracted using a rectangular ERB filterbank with $N_{erb}$ bands applied to the normalized log-power spectrogram. Complex features, $X_{df}(k,f')$, are obtained by extracting the first $N_{feat}$ frequency bands up to the frequency $f_{df}$ from the complex spectrogram and applying band-wise unit normalization.

An ERB decoder then converts this information into predicted gains for each ERB band: $G_{erb, i}(k,b)$ and $G_{erb, s}(k,b)$, corresponding to the impulsive and stationary background components, respectively. An inverse ERB filterbank is applied to these gains, which are then used to filter the input spectrogram, yielding partially extracted impulsive $Y_{G,i}(k,f)$ and stationary $Y_{G,s}(k,f)$ spectrograms. This first filtering stage provides an initial coarse processing, which is then refined by the deep filtering step.

A DF decoder predicts two complex filters, $C_i^M$ and $C_s^M$, applied up to the $N_{feat}$-th frequency band of the spectrograms obtained after the ERB stage. These filters separate the impulsive and stationary spectrograms $\hat{Y}_i(k,f)$ and $\hat{Y}_s(k,f)$:
\begin{align}
    &\hat{Y}_i(k,f') = \sum_{m=0}^M C_i^M (k,m,f') \cdot Y_{G,i}(k-m, f'), \\
    &\hat{Y}_s(k,f') = \sum_{m=0}^M C_s^M (k,m,f') \cdot Y_{G,s}(k-m, f'),
\end{align}
where $M$ denotes the order of the complex filter. This deep filtering stage is performed up to $f_{df} \approx 6$ kHz, where most of the background spectral content resides and where the blending of impulsive and stationary components occurs. Above $f_{df}$, only real-valued filtering is applied. This two-step approach reduces both memory and computational costs by minimizing the size of the complex filters required for source separation.

For further details on the model architecture we refer to the DeepFilterNet2 article \cite{schroter2022deepfilternet2}, which is reproduced as is, with the only difference that the decoders' outputs are doubled for the prediction of each source as shown in \autoref{fig:system_overview}.

\subsection{Loss functions}
We adopt the same loss functions as described in \cite{schroter2022deepfilternet2} for reconstructing each source, i.e., the impulsive and stationary background components, as well as for the mixture $\hat{Y}_i + \hat{Y}_s$.
For each source $Z$ to reconstruct, we compute the following spectrogram loss ($\mathcal{L}_{SP}$) between the predicted source $\hat{Z}$ and the target $Z$,
\begin{equation}
    \mathcal{L}_{SP}(\hat{Z}, Z) = \Vert \vert \hat{Z} \vert^{c} - \vert Z \vert^{c} \Vert_2^2 + \Vert \vert \hat{Z} \vert^{c} e^{j \Phi_{\hat{Z}}} - \vert Z \vert^{c} e^{j \Phi_Z} \Vert_2^2,
\end{equation}
where $\Vert \cdot \Vert_2$ denotes the $l_2$-norm, $c = 0.6$ is a compression factor used to approximate perceived loudness \cite{valin2021low} and $\Phi$ represents the phase.
Additionally, a multi-resolution (MR) spectrogram loss is computed by converting $Z$ back to the time domain using an inverse short-time fourier transform (ISTFT), followed by multiple STFTs with different window sizes of length $\{6, 12, 23\}$ ms:
\begin{multline}
    \mathcal{L}_{MR}(\hat{Z}, Z) =\sum_{m} \Vert \vert \hat{Z}_m \vert^{c} - \vert Z_m \vert^{c} \Vert_2^2 \\ + \sum_{m} \Vert \vert \hat{Z}_m \vert^{c} e^{j \Phi_{\hat{Z}}} - \vert Z_m \vert^{c} e^{j \Phi_{Z}} \Vert_2^2,
\end{multline}
where $m$ indexes the window sizes. The overall loss for the element $Z$ is then given by
\begin{equation}
    \mathcal{L}(\hat{Z}, Z) = \lambda_{SP} \mathcal{L}_{SP}(\hat{Z}, Z) + \lambda_{MR} \mathcal{L}_{MR}(\hat{Z}, Z),
\end{equation}
with $\lambda_{SP} = 1000$ and $\lambda_{MR} = 500$. Finally, the complete training loss sums the contributions of each source to be reconstructed and the mixture:
\begin{equation}
    \mathcal{L}_{total} = \lambda_i \mathcal{L}(\hat{Y}_i, Y_i) + \lambda_s \mathcal{L}(\hat{Y}_s, Y_s) + \lambda_m \mathcal{L}(\hat{Y}_i+\hat{Y}_s, X).
\end{equation}
where $\lambda_i = \lambda_m = 1$ and $\lambda_s = 10$. Since most of the stationary signal is easy to reconstruct—impulse-free regions require little to no filtering—the loss $\mathcal{L}(\hat{Y}_s, Y_s)$ tends to be dominated by these areas. Early experiments showed that applying a weight $\lambda_s > 1$ improves the reconstruction of short segments around impulses by giving them more weight. As for the loss on the mixture, this is calculated to ensure a certain cohesion between the predictions of the background and impulsive sounds, but it remains globally dominated by the other two terms.
\section{Data Generation Pipeline}
\label{section:data}

As mentioned earlier, the key step in the proposed system is the data generation process. We generated training, validation, and test datasets by combining stationary background acoustic scenes with clean impulsive sounds to replicate realistic acoustic environments. The background datasets selected are Dcase2018 Task 1 \cite{Mesaros2018_DCASE}, Cas2023 \cite{bai2024description}, CochlScene \cite{jeong2022cochlscene}, LitisRouen \cite{rakotomamonjy2014histogram}, and ARTE \cite{weisser2019ambisonic}, which provide a wide variety of acoustic scenes and are commonly used in acoustic scene classification tasks. Additionally, we generated synthetic background scenes by augmenting pink noise with random equalization, gain transitions, reverberation, and the addition of low-level Gaussian noise to simulate stationary noises, such as ventilation noise.

For impulsive sounds, we used datasets containing isolated sound events: ESC50 \cite{piczak2015esc}, Nonspeech7k \cite{rashid2023nonspeech7k}, ReaLISED \cite{mohino2022introducing}, and VocalSound \cite{gong2022vocalsound}. We also included two datasets of one-shot percussive instruments: FreesoundOneShotPercussive \cite{ramires2020} and other drum samples. To further increase the variety of impulsive sounds, we generated synthetic events from chirps, harmonic summation, and AR filtering of white noise modulated by asymmetric Gaussian envelopes.
All the code for generating the synthetic sounds, both backgrounds and impulses, will be made publicly available.

\subsection{Impulsive sounds}
In this work, we define an \textit{impulsive acoustic event} as a brief and isolated sound that perceptually stands out from the ambient background. An important aspect of this definition is the fact that the superposition or repetition of impulsive sounds over time, which form a distinct sound layer (e.g., applause or rain), are not considered as impulsive acoustic events. For example, we differentiate between an isolated hammer blow (impulsive) and a continuous burst of jackhammer blows over several seconds (texture to remain in the background component of our model).
\subsection{Dataset Pre-processing}
\begin{figure}[ht]
    \centering
    \includegraphics[width=0.9\linewidth]{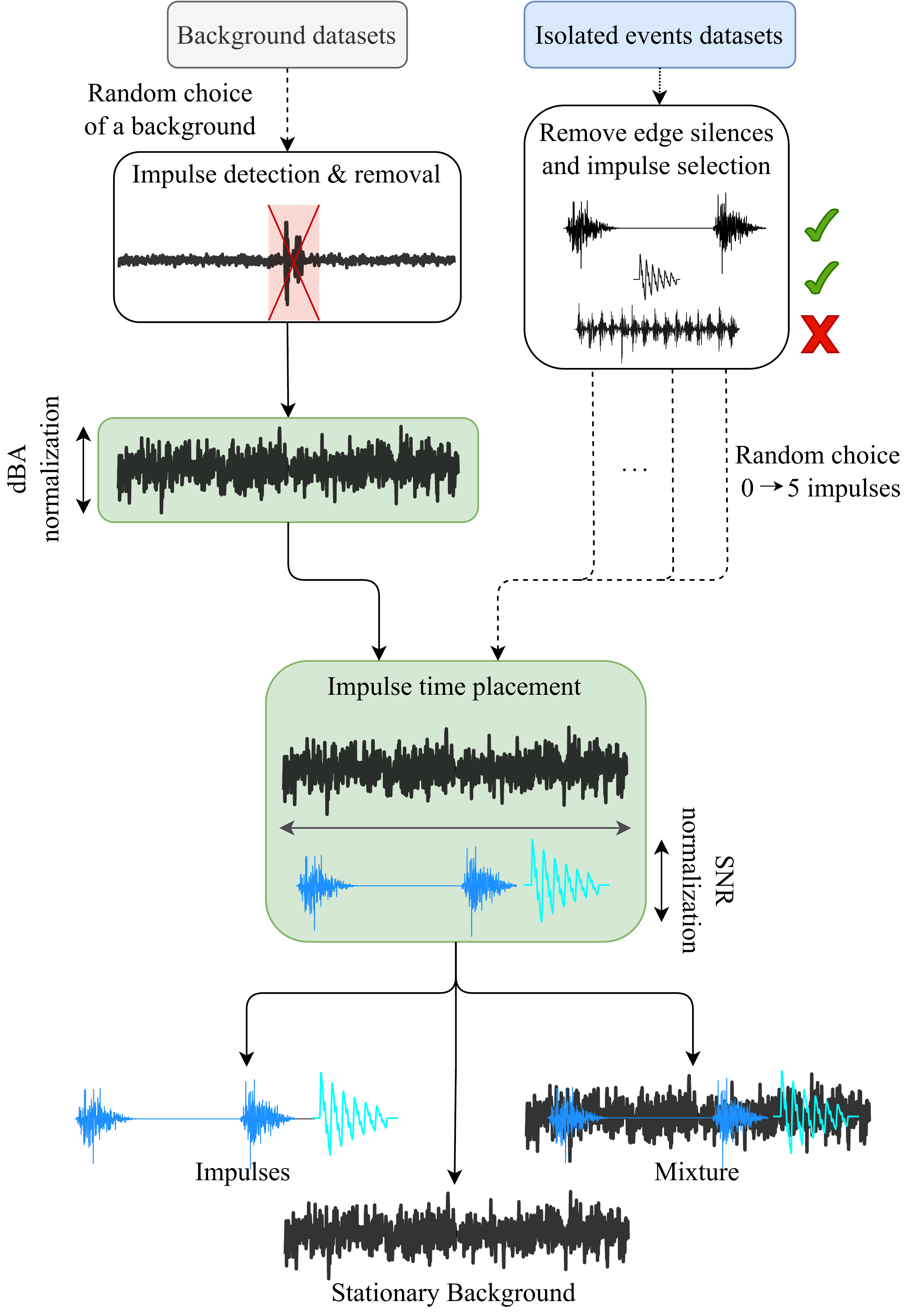}
    \caption{Data generation pipeline.}
    \label{fig:data_generation}
\end{figure}
To generate suitable training data, it is essential to have \textit{clean} background datasets (i.e.~free from discernible impulsive sounds) and clean sounds that are genuinely impulsive. Consequently, all datasets require pre-processing to remove unwanted elements.

For background datasets, we apply an impulse removal procedure at a reduced sampling rate of 16 kHz. First, we detect onsets using \texttt{librosa} with a hop length of 512 samples and a delta threshold of 20\% of the signal's maximum amplitude. Then, a Gabor decomposition is applied to a 5-second window around each detected onset using multi-Gabor dictionaries \cite{pruuvsa2021fast} with various temporal supports ($N_w={32, 64, 128, 512}$ ms).
to obtain the atoms coefficients $c_w(n_w,f_w)$, $n_w$ and $f_w$ being the time and frequency bins. An impulsive event is characterized by a localized burst of energy across frequencies. To identify this pattern in the coefficients obtained, we sum the frequency contributions at each time step $\tilde{c}(n_w) = \sum_{f_w} c_w(n_w, f_w)$. For each window size, the coefficients are interpolated to the resolution of the smallest window, and the contributions from the 3 smallest windows are summed $\tilde{c}_{overall} = \tilde{c}_{w_1} + \tilde{c}_{w_2} + \tilde{c}_{w_3}$. A peak detection is performed on the obtained coefficients with a distance parameter of 100 ms, a prominence of 30\% the maximum value of $\tilde{c}_{overall}$ and a height parameter equal to the maximum coefficient value $c_{w_4}$ on the larger temporal support. If a detected peak is within 200 ms of the onset, the onset is validated as an impulsive sound and removed, with crossfading applied to reconnect the background segments.

For datasets containing isolated sound events, we filter out non-impulsive sounds by applying the following procedure: for each audio sample in the dataset, we first remove the silences at the edges; then, we retain only the sounds shorter than half a second. For longer sounds, we calculate the proportion of silence from the root mean square (RMS) envelope (using a 5\% threshold of the 99th percentile of the envelope) and keep only those with a sufficiently high ratio of silence (50\% for signals less than 1s, and 75\% for longer ones).

\subsection{Dataset generation}

The generated dataset consists of 5-second acoustic scenes, sampled at $f_s = 44100$ Hz, each comprising a background selected randomly from the pre-processed background datasets and several impulsive events chosen from the pre-processed isolated sound datasets. The generation process is presented \autoref{fig:data_generation}.

\begin{figure*}
    \centering
    \includegraphics[width=0.95\textwidth]{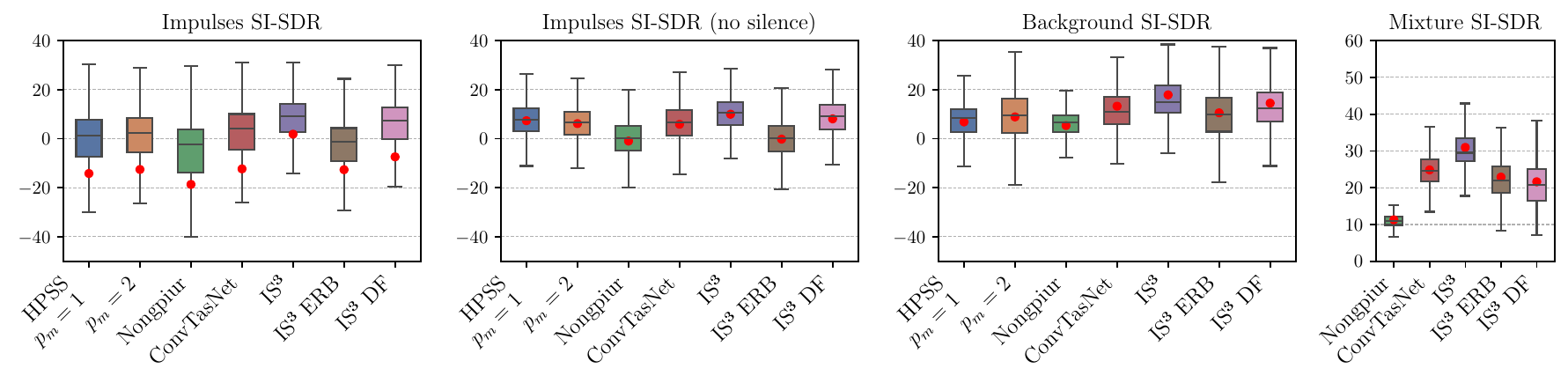}
    \caption{Obtained SI-SDR for the impulses, the stationary backgrounds and the mixture, with the different baseline configurations and the proposed IS³ system. The red dots are the mean values.}
    \label{fig:results}
\end{figure*}

To prevent producing a biased dataset and to avoid over- or under-representation of different acoustic scene categories and impulsive sound types, we organize and unify the various dataset labels, impulse events and backgrounds separately, into a taxonomy using the SALT framework \cite{stamatiadis2024saltstandardizedaudioevent}. When drawing background or impulsive sounds, they are selected from subsets of the datasets that contain the same number of samples for each class, with the exception of those with very few items.
The background track, $y_b(t)$, is normalized to a dBA level sampled from a realistic distribution based on the scene label, while the impulsive sounds are placed randomly without overlapping along the time axis, normalized to reach randomly selected target signal-to-noise ration (SNR) levels, and grouped into a single track, $y_i(t)$. Augmentations are applied to the impulsive sounds (e.g., equalization, reverb, time stretching, and pitch shifting), and a final impulse response is applied to both the background and impulsive tracks. Each set of sources is then exported, as well as the mixture.
In total, 50 hours of mixture data were generated for training, 20 hours for validation, and 10 hours for testing.

\section{Evaluation}
We present here the evaluation framework used to assess the performance of our proposed method, detailing the training setup, baseline methods and the obtained results on the test set.
\subsection{Training setup and baselines}
%\textcolor{blue}{à voir, tout est décrit avant, juste re-préciser batch size, lr dans les résultats}
IS³ was trained with the 50h dataset described in Section \ref{section:data} for a maximum of 150 epochs with early stopping using a batch size of 32 and an Adam optimizer \cite{kingma2014adam} with a learning rate of $10^{-3}$. The audio parameters are the following: we use $N_{erb} = 24$ and $N_{feat} = 256$, just under 6 kHz, and an order $M=8$ for the complex filter. The result is a model with around 2.2 million parameters. Additionally we trained ERB-only (1.2 M parameters) and deep-filtering-only (1.4 M parameters with the same $N_{feat}$) variants of IS³ as ablation studies to demonstrate the benefits of the staged design.

We compare our approach with several baselines. First, we reimplement Nongpiur's method \cite{nongpiur2008impulse} for impulsive noise removal in speech signals, which employs Daubechies wavelets to decompose the noisy signal. The method identifies impulses by applying median filtering to wavelet coefficients across scales and attenuates them to neighboring median levels. We adapt this method to our use case and sampling rate by modifying the wavelet order—choosing 13 instead of 6—and adjusting the factors that control the dynamic thresholds for detecting impulse coefficients: $k_s = 2$ for fine scales and $k_c = 1$ for coarser scales. Additionally, while the original approach only predicts a clean signal with attenuated impulses, we extend it to include an impulsive sound reconstruction stage by retaining only the wavelet coefficients identified as impulses.
Secondly, we compare with the median-filtering HPSS masking method \cite{fitzgerald2010harmonic}, specifically its \texttt{librosa} implementation \cite{mcfee2015librosa, driedger2014extending}. In this baseline, the percussive component is treated as the impulsive part, while the harmonic and residual components (if the margin parameter is greater than 1 are considered as the stationary background. We assess different HPSS configurations by varying the margin parameter, or separation factor, $p_m$. Finally, we re-trained a Conv-TasNet model \cite{luo2019conv}, using \texttt{asteroid} \cite{Pariente2020Asteroid}, adapted to 44.1 kHz input \cite{defossez2019music} (6.3M parameters) to compare with a time-domain neural architecture.
All the code for the data generation, the baselines as well as IS³ are available on github\footnote{https://github.com/ClementineBerger/IS3}.

\subsection{Results and discussion}
The evaluation was conducted using the previously described test set. We use the scale-invariant signal-distortion-ratio (SI-SDR) metric \cite{le2019sdr} on the separated impulsive and stationary background components, as well as on the reconstructed mixtures for Nongpiur's method and IS³. Since HPSS is a masking-based method, its mixture reconstruction is inherently perfect. Additionally, for impulsive sounds, the SI-SDR is computed both with and without silent segments to assess not only reconstruction quality but also the model's ability to preserve silences and prevent background leakage. Statistical significance between IS³ and each baseline configuration is evaluated using the Wilcoxon test over 100 batches of 50 samples, with Bonferroni correction. Results are shown in \autoref{fig:results} and the $p$-values are presented in \autoref{tab:table1}.

\begin{table}[t]
\centering
\caption{Statistical significance between the SI-SDR distributions obtained with the proposed IS³ model and the different baselines and ablations. $p$-values are evaluated using the Wilcoxon test over 100 batches of 50 samples, with Bonferroni correction.}
\label{tab:table1}
\sisetup{
    reset-text-series = false,
    text-series-to-math = true,
    mode=text,
    tight-spacing=true,
    round-mode=places,
    round-precision=2,
    table-format=2.2,
    table-number-alignment=center
}
%\begin{tabular}{l*{2}{S[round-precision=1,table-format=2.1]S}}
\begin{tabular}{l@{\hskip 4pt}*{2}{S[round-precision=1,table-format=2.1]@{\hskip 4pt}S}}
    \toprule
    & \multicolumn{2}{c}{\textbf{Impulse $p$-value}} & \textbf{Background} & \textbf{Mixture}\\
    \cmidrule(lr){2-3}
    & {Silence} & {No silence} & & \\
    \midrule
    {\scriptsize \shortstack{HPSS\\$p_m=1$}} & {$3.74 \cdot 10^{-2}$} & {$0.17$} & {$1.40\cdot 10^{-5}$} & {--} \\
    {\scriptsize \shortstack{$p_m=2$}} & {$5.27 \cdot 10^{-2}$} & {$6.02 \cdot 10^{-2}$} & {$2.66 \cdot 10^{-3}$} & {--} \\
    {\scriptsize Nongpiur} & {$4.96 \cdot 10^{-3}$} & {$1.47 \cdot 10^{-5}$} & {$1.19 \cdot 10^{-7}$} & {$5.18 \cdot 10^{-10}$}\\
    {\scriptsize Conv-TasNet} & {$1.75 \cdot 10^{-8}$} & {$6.72 \cdot 10^{-8}$} & {$8.62 \cdot 10^{-7}$} & {$1.60 \cdot 10^{-7}$} \\
    {\scriptsize IS³ ERB} & {$2.08 \cdot 10^{-8}$} & {$2.26 \cdot 10^{-9}$} & {$5.44 \cdot 10^{-9}$} & {$5.93 \cdot 10^{-8}$} \\
    {\scriptsize IS³ DF} & {$3.96 \cdot 10^{-7}$} & {$8.08 \cdot 10^{-6}$} & {$2.06 \cdot 10^{-5}$} & {$6.62 \cdot 10^{-9}$} \\
    \bottomrule
\end{tabular}
\end{table}

IS³ consistently achieves higher SI-SDR scores for both impulses and backgrounds. The small gap between SI-SDR scores with and without silence suggests that our method accurately reconstructs both impulses and interleaving silences. 
In contrast, HPSS and Nongpiur's method suffer significant SI-SDR degradation when silences are considered, indicating background leakage into the impulsive sound track. This leakage also lowers HPSS’s background reconstruction performance compared to IS³. Finally, while our approach does not strictly guarantee perfect mixture reconstruction like masking methods, it achieves a remarkably high SI-SDR on the mixture. 
Conv-TasNet and ablations achieve intermediate results demonstrating the value of the architecture chosen for IS³.

Finally, it is important to note that both signal processing baseline methods suffer from a reliance on a challenging and highly impulsive noise type dependant parameter selection. This dependency reduces their performance in our experiments, which involve a wide variety of impulsive sound types. In contrast, our approach offers superior generalization and eliminates the need for noise-specific parameter tuning. For a qualitative comparison, audio examples are provided in the supplementary materials\footnote{https://clementineberger.github.io/IS3/} for both synthetic and real-world data. 
%Those examples show that classic signal processing methods suffer from notable leakage between sources, whereas IS³ achieves clearer separation with minimal leakage. That said, we observed some slight “vacuum-like” artifacts in stationary backgrounds when impulsive events are removed—though crucially, not resulting in complete silencing.

%\textcolor{blue}{il me semble qu'il existe des versions améliorées avec des réseaux de neurones de l'algo HPSS, mais ça me semble peu pertinent ici vu qu'ils sont entrainés sur de la musique et donc encore plus spécialisés là-dedans que le HPSS basique. à préciser pour justifier qu'on ne s'intéresse qu'à ça comme baseline ?}

% dire que c'est un peu biaisé car l'évalution est réalisée sur un dataset synthétique créé selon nos critères en termes de distinction entre background stationaire et évènements isolés. Notre modèle est entrainé pour traiter spécifiquement ce type de sons. La ligne où se situe la distinction entre son impulsif et son stationnaire est moins évidente pour la méthode HPSS -> d'où des audios réels à écouter sur la page compagnon.
\section{Conclusions}

In this paper, we have introduced IS³, a solution for Impulsive--Stationary Sound Separation, designed to isolate generic impulsive acoustic events from stationary backgrounds within an acoustic scene.
Our approach leverages and adapts the DeepFilterNet2 two-stage deep filtering process for this task and is trained using a dedicated dataset generated through a sophisticated data generation pipeline to ensure diversity and balance across sound categories. Evaluation results demonstrate that IS³ is successful at separating both impulsive and stationary components while minimizing background leakage. These results demonstrate that a learning-based approach trained on well-designed data is well-suited for the task and can achieve strong performance even with a relatively lightweight model.
In particular, the proposed approach outperforms the classic HPSS masking method and wavelet filtering by a large margin in terms of SI-SDR.

% The preferred spelling of the word acknowledgment in America is without an ``e'' after the ``g.'' Try to avoid the stilted expression, ``One of us (R.\ B.\ G.) thanks ...'' Instead, try ``R.\ B.\ G.\ thanks ...''  Put sponsor acknowledgments in the unnumbered footnote on the first page. Please include acknowledgments only in the camera-ready version, and NOT in the version of the paper submitted for review.

% -------------------------------------------------------------------------
% Either list references using the bibliography style file IEEEtran.bst

\clearpage
% The \IEEEtriggeratref{XX} command can be used to move to the next column before the XX-th reference
% to balance the two columns of the reference section
% \IEEEtriggeratref{XX}
\bibliographystyle{IEEEtran}
\bibliography{refs25}
% or list them by yourself:
% \begin{thebibliography}{1}

% \bibitem{waspaaweb}
% {WASPAA Website}, \url{http://www.waspaa.com}.

% \bibitem{IEEEXploreReqs}
% {IEEE {X}plore {R}equirements}, \url{https://conferences.ieeeauthorcenter.ieee.org/write-your-paper/meet-ieee-xplore-requirements/}.

% \bibitem{eWilliams1999}
% E.~Williams, \emph{Fourier Acoustics: Sound Radiation and Nearfield Acoustic Holography}.\hskip 1em plus 0.5em minus 0.4em\relax London, UK: Academic Press, 1999.

% \bibitem{cJones2003}
% C.~Jones, A.~Smith, and E.~Roberts, ``A sample paper in conference proceedings,'' in \emph{Proc. ICASSP}, vol.~II, Apr. 2003, pp. 803--806.

% \bibitem{aSmith2000}
% A.~Smith, C.~Jones, and E.~Roberts, ``A sample paper in journals,'' \emph{IEEE Trans. Signal Process.}, vol.~62, pp. 291--294, Jan. 2000.

% \end{thebibliography}

\end{document}